\documentclass{article}

\usepackage{arxiv}
\usepackage{amsmath}
\usepackage{amssymb}
\usepackage{amsfonts}
\usepackage{mathtools}
\usepackage[most]{tcolorbox}
\usepackage{courier}
\usepackage{xurl}
\usepackage[utf8]{inputenc} 
\usepackage[T1]{fontenc}    
\usepackage{hyperref}       
\usepackage{url}            
\usepackage{makecell} 
\usepackage{booktabs}
\usepackage{pifont}

\usepackage{placeins}
\usepackage[authoryear]{natbib}
\usepackage[table]{xcolor}

\usepackage{booktabs}       
\usepackage{amsfonts}       
\usepackage{nicefrac}       
\usepackage{microtype}      
\usepackage{tikz}
\usepackage{pgfplots}\pgfplotsset{compat=1.17}
\usetikzlibrary{calc,arrows.meta}
\usepackage{lipsum}
\usepackage{graphicx}
\graphicspath{ {./images/} }
\usepackage[utf8]{inputenc} 
\usepackage[T1]{fontenc}    
\usepackage{hyperref}       
\usepackage{url}            
\usepackage{booktabs}       
\usepackage{amsfonts}       
\usepackage{nicefrac}       
\usepackage{enumitem}
\usepackage{microtype}      
\usepackage{xcolor}         
\usepackage{wrapfig}
\usepackage{graphicx}
\usepackage{listings}
\usepackage{subcaption}
\usepackage{float}  
\title{Predict future sale}
\usepackage{ragged2e}

\definecolor{deeppurple}{HTML}{4A1D6E}
\definecolor{ourstint}{HTML}{EDE6F5}
\definecolor{yesgreen}{HTML}{2E7D5B}
\definecolor{nogray}{HTML}{B0A8BB}
\newcommand{\yes}{\textcolor{yesgreen}{\checkmark}}
\newcommand{\no}{\textcolor{nogray}{$\times$}}
\newcolumntype{L}{>{\RaggedRight\arraybackslash}m{0.22\textwidth}}
\newcolumntype{M}{>{\RaggedRight\arraybackslash}p{0.52\textwidth}}
\newcolumntype{Y}{>{\centering\arraybackslash}m{0.16\textwidth}}
\usepackage[normalem]{ulem}   
\newif\ifcomments
\commentstrue
\ifcomments
  
  \newcommand{\rb}[1]{\textcolor{magenta}{[rishi: #1]}}
  \newcommand{\stephane}[1]{\textcolor{blue}{[Stephane: #1]}}
  \newcommand{\tomas}[1]{\textcolor{green!50!black}{[tomas: #1]}}
  \newcommand{\alex}[1]{\textcolor{purple}{[alex: #1]}}
  \newcommand{\todo}[1]{\textcolor{red}{[TODO: #1]}}
\else
  
  \newcommand{\rb}[1]{}
  \newcommand{\stephane}[1]{}
  \newcommand{\tomas}[1]{}
  \newcommand{\alex}[1]{}
  \newcommand{\todo}[1]{}
\fi

\author{
 Ziyue Qi \\
  School of Coumputing and Information\\
  University of Pittsburgh\\
  Pittsburgh, PA 15213 \\
  \texttt{ziq2@pitt.edu} \\
   \And
 Zixuan Lu \\
  School of Coumputing and Information\\
  University of Pittsburgh\\
  Pittsburgh, PA 15213 \\
  \texttt{ZIL50@pitt.edu} \\
  \And
 Yuchen Lu \\
  School of Coumputing and Information\\
  University of Pittsburgh\\
  Pittsburgh, PA 15213 \\
  \texttt{yul217@pitt.edu} \\
}

\begin{document}

\title{Estimating time spent on work tasks}

%

\author{%
  Stephane Hatgis-Kessell \\
  Stanford University \\
  \And
  Tomás Aguirre \\
  University of São Paulo \\
  \AND
  Alexander Wan \\
  Stanford University \\
  \And
  Rishi Bommasani \\
  Stanford University \\
}


\maketitle

\begin{abstract}
The task-based framework in economics models occupations as bundles of tasks. 
It is the standard lens for understanding how technology affects work: a new technology changes the cost or time each task requires and these task-level effects aggregate to occupation-level effects. 
We study how tasks should be weighted in this aggregation. 
Prior work has relied on idiosyncratic or ill-justified choices for task weights.
While recent work suggests weighting tasks by time spent, existing time shares are either based on coarse O*NET data not intended for this purpose and/or estimated via black-box language models.
We aim to close this gap by proposing a principled method for estimating time shares for nearly 18,000 tasks that constitute nearly all U.S. occupations.
Our estimates factor a task's time into (i) the expected frequency of the task, derived from O*NET, and (ii) the time to complete a single instance of it. 
To estimate the latter quantity, we solve a constraint satisfaction problem based on pairwise comparisons elicited from language models about which tasks are longer per instance.
We validate our estimates by characterizing the solution space of the constraint satisfaction problem and collecting data from workers for multiple occupations.
We apply our time shares to analyze how AI exposes U.S. occupations and find that some prior results are sensitive to time weights. 
Accounting for the share of working time exposed to AI, rather than the share of tasks like prior work, widens the gap between the least and most exposed occupations: it lowers measured exposure for most occupations but raises it for the most exposed. 
Re-weighting by time also reshuffles 11 of the 25 occupations widely reported as most exposed to AI, shifting the top of the list away from clerical work and toward analytical roles. 
Time shares can serve as a general primitive for research and policy on the labor economy and the economics of technology.
\end{abstract}

\section{Introduction}
\citet{autor2003skill} popularized the task-based approach in economics in which occupations are modeled as bundles of tasks.
Subsequently, many works estimate occupation-level outcomes — employment, wages, skill demand, automation exposure, technological displacement — by aggregating across the occupation's constituent tasks. This approach is powerful because tasks provide a structured way to reason about an occupation and pinpoint where a technology acts: by operating at the task level, one can model how a tool reshapes an occupation through the specific tasks it affects, rather than treating the occupation as an indivisible whole.
For example, \citet{autor2003skill} show that the task content of jobs shifted substantially over three decades even as job titles and employment shares held steady — change that occupation-level analysis would largely miss. More recently, task-based frameworks have been widely adopted to study the potential impacts of AI on labor markets \citep{webb2020impact, felten2021occupational, brynjolfsson2018what, eloundou2024gpts}. \citet{autor2025expertise} also show that changes in the expertise level of tasks that constitute a job predict occupational wage and employment changes.

In this paper, we study the aggregation function: how should tasks be weighted to compute occupation-level quantities? 
Many works that adopt the task-based framework confront this question but there is no established consensus on how to aggregate.
Since tasks are not equally important, most works use some set of weights based on available task-level metadata. 
However, the conceptual basis for using these heuristics to aggregate is often unexplained. 

A natural alternative is to weight tasks proportional to the time workers spend on each task on average.
Economic models in several domains depend on such time weights \citep{bouquet2026intensive, hosseini2026generative_entry_barriers, martin2022green, tamkinmccrory2025productivity, battisti2023technological, peri2009task}, but lack empirically grounded values and instead rely on heuristics that may diverge from actual time spent. Time spent is not only useful for assigning weights in a task-based model, but also a fundamental characterization of work.
Wages are denominated per unit time: some professions (e.g., lawyers, consultants) directly bill for labor costs as a function of time spent.
More broadly, time spent on work and the tasks therein is central to broader accounting of how people allocate their time.
The American Time Use Survey (ATUS) has been administered for over two decades to understand how Americans spend their day.\footnote{The activities reported in the ATUS are insufficiently granular to estimate time spent on individual work tasks.}

We address the gap by providing time spent estimates for every occupation across all of its tasks.
To map from occupations to tasks, we rely upon the U.S. Department of Labor's O*NET taxonomy, which decomposes 1,016 occupations into 18,796 tasks. We design our estimation method to make our assumptions about work explicit, rather than delegating estimation of the key quantity to a language model (LM). We factor the total time spent on a task into two components: the expected task frequency (e.g., attending 5 meetings per week) and the time spent on an individual \textit{instance} of a task (e.g., spending 30 minutes on a meeting). The expected task frequency is computed from O*NET incumbent surveys and U.S. labor data.
To estimate the time spent on an individual instance of a task, we specify a set of constraints and solve the resulting constraint satisfaction problem using linear programming.
The first constraint is a per-occupation daily-time budget: the frequency a task is performed times the duration of a single instance of that task (the object we are computing), summed across all tasks belonging to an occupation, must equal a typical workday. 
The second is a set of ordinal constraints on single-instance durations — which of two tasks takes longer per instance — elicited from an LM via pairwise comparisons, on the premise that LMs can rank tasks by which ones take longer for a typical worker to complete. 

Time encodes how technology changes work.
For example, new technologies can decrease time spent on tasks to confer productivity gains to workers, firms, and the overall economy.
Recently, several works have considered how AI can reduce the time spent on tasks through ex-ante exposure estimates \citep[e.g.,][]{eloundou2024gpts,  hosseini2026generative_entry_barriers} and ex-post micro studies \citep[e.g.,][]{noy2023experimental, peng2023impact}.
For example, the exposure estimates from the method of \citet{eloundou2024gpts} indicate that Massage Therapists are highly exposed to language model capabilities because 44\% of the tasks associated with the occupation are exposed.
These tasks where language models could be most useful include developing client treatment plans, maintaining treatment records, and referring clients to other therapists.
But our estimates of time spent per task indicate these tasks only constitute 16\% of the work hours of Massage Therapists, which results in Massage Therapists no longer being highly exposed (see Figure \ref{fig:massage-exposure}).
Incorporating time share enables a quantity not captured by existing measures — the fraction of working time, rather than the fraction of tasks, exposed to LMs — which is arguably an object of interest for questions about labor reallocation and displacement

\definecolor{exposed}{HTML}{C44536}    
\definecolor{exposedbg}{HTML}{F6E0DB}  
\definecolor{timecol}{HTML}{2F5D62}    

\newcommand{\exrow}[2]{\rowcolor{exposedbg}\textbf{#1} & \textbf{#2}\\}

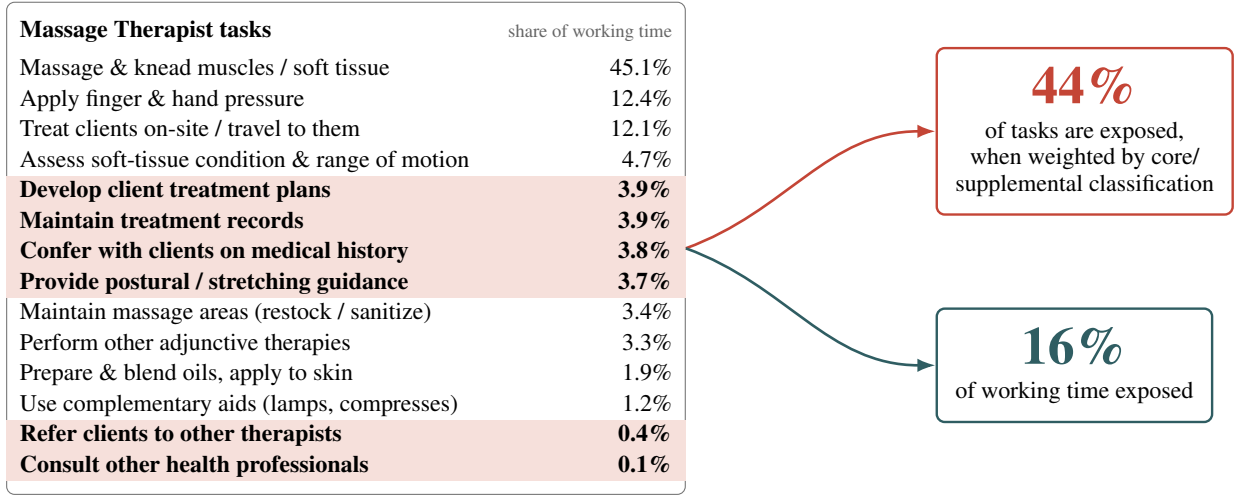
\begin{figure}[t]
\centering
\begin{tikzpicture}[
    every node/.style={font=\footnotesize},
    resultbox/.style={draw, line width=0.9pt, rounded corners=3pt,
                      align=center, inner sep=7pt, minimum width=2.9cm},
    flow/.style={-{Latex[length=2.4mm]}, line width=1pt},
  ]

  \node[draw=black!55, rounded corners=3pt, inner sep=5pt, fill=white] (tasks) {%
    \setlength{\tabcolsep}{5pt}\renewcommand{\arraystretch}{1.15}%
    \begin{tabular}{@{}p{6.1cm}r@{}}
      \textbf{Massage Therapist tasks} & {\scriptsize\color{black!60}share of working time}\\[3pt]
      Massage \& knead muscles / soft tissue        & 45.1\%\\
      Apply finger \& hand pressure                 & 12.4\%\\
      Treat clients on-site / travel to them        & 12.1\%\\
      Assess soft-tissue condition \& range of motion & 4.7\%\\
      \exrow{Develop client treatment plans}{3.9\%}
      \exrow{Maintain treatment records}{3.9\%}
      \exrow{Confer with clients on medical history}{3.8\%}
      \exrow{Provide postural / stretching guidance}{3.7\%}
      Maintain massage areas (restock / sanitize)   & 3.4\%\\
      Perform other adjunctive therapies            & 3.3\%\\
      Prepare \& blend oils, apply to skin           & 1.9\%\\
      Use complementary aids (lamps, compresses)    & 1.2\%\\
      \exrow{Refer clients to other therapists}{0.4\%}
      \exrow{Consult other health professionals}{0.1\%}
    \end{tabular}};

  \node[resultbox, draw=exposed, anchor=west]
    (hl) at ($(tasks.east)+(3.3cm,1.55cm)$)
    {\textcolor{exposed}{\textbf{\fontsize{20}{20}\selectfont 44\%}}\\[3pt]
     of tasks are exposed,\\ when weighted by core/\\supplemental classification};
  \node[resultbox, draw=timecol, anchor=west]
    (tw) at ($(tasks.east)+(3.3cm,-1.55cm)$)
    {\textcolor{timecol}{\textbf{\fontsize{20}{20}\selectfont 16\%}}\\[3pt]
     of working time exposed};

  \coordinate (src) at (tasks.east);
  \draw[flow, exposed] (src) to[out=20,in=180] (hl.west);
  \draw[flow, timecol] (src) to[out=-20,in=180] (tw.west);

\end{tikzpicture}

\caption{Constituent tasks for Massage Therapists from O*NET. The left column shows our estimated share of working time per task; tasks measured as exposed to AI by \cite{eloundou2024gpts} are highlighted in red. Under their weighting scheme, where tasks labeled as core to the job are weighted 1 and supplemental tasks 0.5, $44\%$ of the occupation's tasks are exposed. We estimate these tasks occupy only $16\%$ of working time.}
\label{fig:massage-exposure}
\end{figure}

Building on this example, we use our time estimates to demonstrate how prior findings about the impacts of AI on jobs update when made time-sensitive.
Overall, we find that time-sensitivity widens the gap between the least and most exposed occupations.
For occupations with a low-to-moderate share of exposed tasks, the time-weighted exposure share is even lower; whereas for occupations with the highest share of exposed tasks, the share of time exposed is higher still.
Broadly, accounting for time share corrects for occupations with highly exposed tasks that do not take up large shares of worker time (e.g., clerical duties) or highly exposed tasks that do take a lot of time (e.g., analytical, research, or writing tasks). 
Time weights change the headline findings that are most covered by the media about the impact of AI on jobs.

For example, 11 of the 25 occupations labeled as most exposed according to the method of  \citet{eloundou2024gpts}, which have been widely discussed in popular press \citep[e.g.,][]{weber2023wsj, euronews2023openai} and government reports \citep{sanders2025bigtech,scavette2025occupational,gmyrek2024genai}, change when re-weighting tasks by the share of time spent.\footnote{We recompute exposure using the method of \citet{eloundou2024gpts} with the updated prompt of \citet{hosseini2026generative_entry_barriers}, so our top-25 list is comparable to, but not identical to, the one originally reported by \citet{eloundou2024gpts} and cited in popular press and government and policy reports.}
Given that time is a general primitive, we expect that our estimates can be widely adopted in the study of the economics of technology and the labor economy.

We release the code used to produce our results, and time share weights for nearly all tasks in O*NET 30.2.\footnote{\url{https://github.com/Stephanehk/Estimating-Time-Spent-On-Work/tree/main}}

\section{Related Work}
Our work builds upon the task-based framework and could either complement or substitute for other approaches to task-to-occupation aggregation.
We also relate our work to recent approaches in the study of AI's economic impacts that have explored related notions of time spent.

\textbf{The task-based framework.} 
The task-based framework introduced by \citet{autor2003skill}, where occupations are represented as bundles of tasks, has produced now-canonical findings about how technology reshapes work. This includes accounting for much of the polarization of the U.S. labor market \citep{autor2003skill, autor2013growth}. \citet{acemoglu2011skills} and predicting unemployment risk or occupational changes from technology (e.g., \citet{autor2025expertise, frey2017future, frank2023ai, eloundou2024gpts, hosseini2026generative_entry_barriers, handa2025economictasksperformedai}).

The power of this framework comes from the fact that technologies do not act on occupations directly; they act on tasks. A new tool rarely automates or augments an entire job at once—it changes the cost, speed, or feasibility of specific activities within that job, and the occupation-level outcome is an aggregate of those task-level effects. Modeling occupations as indivisible units obscures this mechanism. For example, \citet{autor2003skill} found that the economy-wide shift away from routine tasks and toward nonroutine analytic and interactive ones in the US between 1960 and 1998 occurred overwhelmingly \emph{within} occupations and industries rather than through the reallocation of employment across occupations. An occupation-level analysis would have missed most of this change, since the content of jobs was being reshaped by computerization even where job titles and their employment shares held steady. Task-level analysis has also accurately predicted historical employment and wage trends \citep{autor2025expertise} and has become the standard lens for studying AI exposure: it makes occupation-level consequences a transparent function of task-level changes. 

\textbf{Demand for time weights.} 
To make use of the task-level abstraction for reasoning about occupation-level outcomes, aggregation is necessary.
The simplest approach of treating all tasks equally clearly diverges from the reality of many occupations, where some tasks are considerably more critical than others.
For example, the task of cutting hair is facially more central to the occupation of barber than maintaining records. The economics literature on task-based approaches recognizes this. Many works express a preference for time-based weights but, lacking task-level estimates, fall back on other task attributes as proxies.
\autoref{tab:weighting_methods} summarizes prior approaches.

\begin{table}[t]
\centering
\setlength{\tabcolsep}{8pt}
{\renewcommand{\arraystretch}{1.25}
\resizebox{\columnwidth}{!}{%
\begin{tabular}{@{}l l c@{}}
\arrayrulecolor{deeppurple}\toprule
\textbf{Method} &
\textbf{Task Weights} &
\textbf{Measures time share?} \\
\midrule
\arrayrulecolor{black}
\citet{acemoglu2011skills}        & importance                                              & \no  \\
\citet{brandes2016opening}        & frequency, mapped to time share via constrained LP      & \yes \\
\citet{brynjolfsson2018what}      & importance                                              & \no  \\
\citet{webb2020impact}            & average of frequency, importance, and task relevance    & \no  \\
\citet{felten2021occupational}    & importance $\times$ paper-specific measure              & \no  \\
\citet{martin2022green}           & frequency $\times$ author-assigned time weights         & \yes \\
\citet{eloundou2024gpts}          & task type classification (core vs.\ supplemental)       & \no  \\
\citet{tamkinmccrory2025productivity} & generated by prompting an LLM                       & \yes \\
\citet{bouquet2026intensive}      & frequency, mapped to time and normalized per occupation & \yes \\
\citet{hosseini2026generative_entry_barriers} & frequency $\times$ importance                   & \yes \\
\midrule
\rowcolor{ourstint}
\textbf{Ours} & \textbf{Satisfy constraints from frequency data and LM-labeled time rankings} & \yes \\
\arrayrulecolor{deeppurple}\bottomrule
\arrayrulecolor{black}
\end{tabular}}}
\vspace{4pt}
\caption{\textbf{Aggregation approaches from tasks to occupations.}
We survey prior task-based research to document aggregation methods, including whether they are intended as proxies for time share. Prior work often relies on O*NET metadata as-is; ours is the first approach, to our knowledge, to estimate per-instance task time, the missing ingredient for a full time-share estimate.}
\label{tab:weighting_methods}
\end{table}

Within the task-based literature, a strand of theoretical work offers an alternative approach to time weights \citep{acemoglu2011skills, gans2026oring, hosseini2026generative_entry_barriers,hampole2025ai,liu2025technology,lindenlaub2026beyond,atalay2017evolving}. 
In these works, time spent is derived rather than observed because the focus is on optimal (rather than realistic) time allocations.
Therefore, a cost-minimization problem is solved (often in closed-form) to yield the time spent per task.
This yields a straightforward theoretical mapping from task-level time savings to occupation-level productivity gains. Because these models target optimal rather than realized allocations, and in idealized conditions, they are not intended to recover the time workers actually spend.

\textbf{Alternative heuristics for task-to-occupation aggregation.} 
Prior to our work, the lack of availability of time share estimates has led prior work to use other aggregation weights based on available resources. 
The most common alternative is to use task-level metadata provided by O*NET since almost all empirical task-based papers already make use of O*NET for the task-to-occupation mapping.

Two ubiquitous options involve O*NET importance scores which are assigned by incumbent workers for each occupation's tasks.
Importance scores are published as the worker-average on a 5-point scale: workers are provided minimal guidance when surveyed on how to interpret ``importance''.\footnote{Workers are asked to ``Rate importance of the task for performance of the occupation'' with no further definition of ``importance''.}
Some works directly weight tasks proportional to these importance scores \citep[e.g.,][]{brynjolfsson2018what, acemoglu2011skills}.
O*NET also categorizes tasks as \textit{core} or \textit{supplemental}:  a core task is rated as relevant by at least 67\% of workers and has an importance score of at least 3 out of 5.
Some works assign different weights to core and supplemental tasks; the most well-known work on the task-based approach for AI is \citet{eloundou2024gpts}, which assigns core tasks a weight of 1 and supplemental tasks a weight of 0.50. 
Task importance, and the associated core-supplemental distinction, is related to time share but may be conceptually distinct: tasks that are highly important to an occupation may be performed either frequently or infrequently and may take varying amounts of time.
For Barbers, for example, cleaning and sterilizing instruments and cutting hair both receive high importance ratings, yet barbers likely spend substantially more time on the latter.
Conversely, a professor may spend a substantial share of their time writing emails despite this task being of low importance to their work.

Other work proposes using O*NET task frequency values to weight tasks \citep[e.g.,][]{bouquet2026intensive}. Task frequency need not correspond to time spent on the task. Frequency alone ignores variation in how long a single instance of each task takes (e.g., answering one phone call versus performing data analysis). Our time share measure accounts for task frequency, but further estimates how long a single instance of each task takes. 

\textbf{Recent explorations of time in relation to AI.} \citet{tamkinmccrory2025productivity} propose two measures of task time relevant to ours: (1) the share of time a worker spends on a task, and (2) the length of a single task instance absent AI assistance. We distinguish between a task — a high-level activity as defined in O*NET (e.g., ``Write fiction or nonfiction prose'') — and a task instance, a specific occurrence of performing that task (e.g., writing the third chapter of a particular novel). The first measure, (1), corresponds directly to the time share quantity we estimate. \citet{tamkinmccrory2025productivity} produce this estimate by prompting an LM with a short description of each task and asking it how much time workers in the occupation spend on it. This procedure departs from our desideratum that estimates rest on transparent, stated assumptions, and its time-share estimates are checked only for self-consistency across prompt variants rather than against external ground truth. The second measure — task instance length — is obtained by prompting an LM to estimate how long a user would have taken to perform a given task instance without AI, using real Claude transcripts as input; each transcript and its estimate are then mapped to an O*NET task. While these estimates are validated against human judgments for a subset of task instances, the underlying quantity differs from ours in two important ways. First, a Claude transcript may span multiple O*NET tasks and is restricted to instances a user can perform through a chatbot (e.g., excluding physical work). Second, the time to complete a single task instance (e.g., writing one chapter) is distinct from the share of a worker's time the corresponding task occupies in a typical week (e.g., an author may only write a few hours per day). Our target is the latter.

\cite{freund2026job} also prompt an LM to estimate the share of time spent per cluster of tasks. Their approach operates over 38 task clusters, while our method aims to estimate the time spent per task for the over 19,000 occupation-specific tasks in O*NET. Further, by way of prompting an LM to directly estimate time shares, their method, like other direct-elicitation approaches, does not state its assumptions explicitly. A separate recent line of work uses time as an axis for measuring AI capability rather than as a weight on occupational tasks. \citet{kwa2025metr} estimate the \textit{time horizon} over which AI systems can autonomously complete tasks — the length of task that a system can finish at a given success rate. This quantity is conceptually distinct from ours: it measures the duration of task instances, not the share of a worker's time occupied by a given O*NET task.

\section{Method}
\begin{figure}[t]
    \centering
    \includegraphics[width=0.7\textwidth]{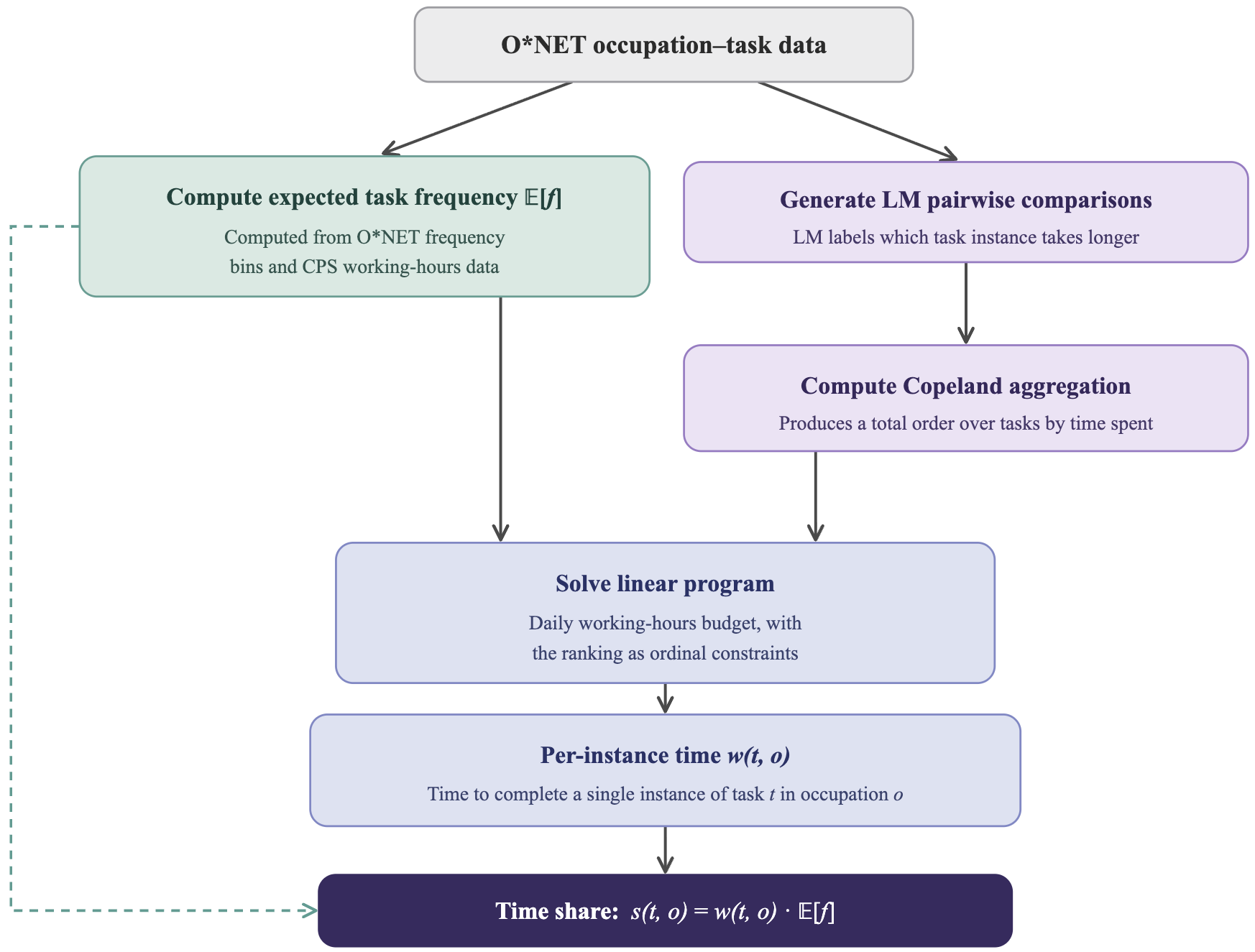}
    \caption{Overview of our method for estimating task time shares for occupations on O*NET. We first compute each task's expected daily frequency: ONET frequency bins are converted into task instances per day, adjusted using CPS data where available, then averaged across all ONET survey respondents. Next, we prompt an LM to rank tasks by which ones take longer to complete a single instance of. The outputs of those two processes are used to generate the constraints for a linear program (\autoref{eq:lp}) which, when solved for, produce time weights. The task time weights, when combined with the task's expected frequency, produce our estimate for share of time spent on the task.}
    \label{fig:method_overview}
\end{figure}

We estimate the share of time spent per task for each occupation's tasks.
We transparently factorize these estimates into (i) the expected frequency of the task and (ii) time per single instance of the task. \citet{bouquet2026intensive} and \citet{brandes2016opening} follow the same factorization, but do not attempt to estimate (ii), the time per single instance of the task, like we do. This factorization relies on the assumption that a task's expected frequency is independent of the time to complete a single instance of it; we further expand on that assumption in Appendix \ref{app:ind_assumption}. Expected task frequency is computed from data published by the U.S. Department of Labor and time per task instance is estimated by solving a linear program grounded in constraints generated by language models (LMs). Our method is summarized in \autoref{fig:method_overview}. 

\paragraph{Occupational data.}
Our estimation procedure assumes (i) a list of occupations and (ii) a mapping from occupations to tasks.
We use the Standard Occupational Classification (SOC) list of 1,016 U.S. occupations and the O*NET taxonomy that maps these 1,016 occupations to 18,796 tasks. In practice, we restrict our method to the 876 occupations for which all required data are available in O*NET.
The data is prepared by the U.S. Department of Labor and updated quarterly to reflect changes to the U.S. labor economy.

\paragraph{Measuring expected task frequency.}
Since 2008, O*NET publishes a task frequency for each worker responding to the occupational survey.
Frequencies are reported in one of seven frequency bins.
We convert these frequencies into units of \textit{task instances per day}.
Define 
\[
F(o) = \left\{\,7,\;2,\;1,\;\tfrac{1}{h_1(o)},\;\tfrac{1}{h_2(o)},\;\tfrac{1}{h_3(o)},\;\tfrac{1}{h_4(o)},\;0\,\right\},
\]
whose entries correspond to the bins hourly or more, several times daily, daily, more than weekly, more than monthly, more than yearly, yearly or less, and never. The terminal never category is imputed from the O*NET relevance field.
For tasks performed less often than daily, we estimate the normalization constants $h_i(o)$ from occupation-level data from the Current Population Survey (CPS) about hours worked per week and weeks worked per year, defaulting to 52 weeks and 40 hours per week when CPS data is unavailable.
Each survey respondent $j$ in occupation $o$ reports a frequency $f_j(t, o) \in F(o)$ of task $t$: let $\mathbb{E}\left[{f}(t, o)\right]$ be the expected daily task frequency. \cite{bouquet2026intensive} follow a similar procedure to map task frequencies to units of hours, but do not adjust for CPS data.

\textbf{Ranking tasks by single-instance time.}
Unlike the task frequencies, O*NET and other public datasets do not report the time spent for each instance of a task.
Therefore, without independently surveying workers or otherwise observing these durations, we rely upon language models to aid in this estimation.
The simplest approach is to directly elicit these values from a language model for every task: this resembles the approach of \citet{tamkinmccrory2025productivity}.
Based on initial experiments, we found these estimates inconsistent with basic constraints. For example, the resulting single-instance time estimates when combined with O*NET task frequencies yielded more time spent on all tasks than would be consistent with the CPS total time worked per week.
Therefore, we formulate an alternative approach that more narrowly constrains how language models are used to generate data and uses traditional methods (i.e., linear programs) to address the remainder.

The key difference is our approach relies on language models to \textit{rank} tasks for which ones take longer for a single instance, rather than fully estimating the time per instance.
We prompt a language model to decide which task takes longer for a single instance for all task pairs.
The model\footnote{We use GPT 5.2 throughout this work.} selects from one of four categories: \emph{Task 1 takes longer}, \emph{Task 2 takes longer}, \emph{Can't tell}, or \emph{Same}. 
We discard pairs labeled \emph{Can't tell} or \emph{Same}, and additionally discard pairs whose label tokens are generated with probability below $70\%$ to filter out low-confidence comparisons.
The exact prompting details are specified in \autoref{fig:time-preference-prompt}.
Given these binary preferences, which may not be transitive and/or may lack preferences for some pairs, we generate a total order (i.e., ranking) over all tasks using the Copeland method from social choice theory.\footnote{Any ranked-choice voting rule can be used: we use the Copeland method as a convenient Condorcet method.}

\textbf{Estimating time spent via linear programs.}
We interpret the ranking of tasks based on the time spent on a single instance as strict constraints.
Combined with the task frequencies, we define a constraint satisfaction problem to find the total time spent on each task.
We introduce one additional constraint, which is the overall occupation-level daily time budget: how much time does a worker spend on average working per day?
Since we completely allocate this budget across all O*NET tasks (assuming no two tasks can be done simultaneously), we set the budget as 7 hours per day to account for workers spending time on activities beyond O*NET tasks (e.g., unrecognized tasks, commuting, lunch, breaks) relative to the standard 8-hour work day.

We define the linear program below.
$w(t, o)$ is the per-instance time for task $t$, $d_o$ is the gap between the task with the most single-instance time $w(t^{\text{top}}_o, o)$ and least $w(t^{\text{bot}}_o, o)$ and $\epsilon$ is the minimum-separation parameter for the margin in the single-instance time per task between consecutive tasks in the ranking.
\begin{equation}
\label{eq:lp}
\begin{aligned}
\max_{\{w(t, o)\},\, \{d_o\}} \quad & \sum_{o} d_o \\
\text{s.t.} \quad
& \sum_{t \,\in\, \mathrm{tasks}(o)} w(t, o)\, \mathbb{E}[f(t, o)] \;=\; 7,
&& \forall\, o, \\
& w(t_i , o) \;\ge\; w(t_j, o) + \epsilon,
&& \forall\, (t_i, t_j) \in \mathcal{R}_o,\; \forall\, o, \\
& d_o \;\le\; w(t^{\text{top}}_o,  o) - w(t^{\text{bot}}_o,  o),
&& \forall\, o, \\
& w(t, o) \;\ge\; 0,
&& \forall\, t, o.
\end{aligned}
\end{equation}

A core design choice is how we handle the possibility of multiple feasible solutions.
The linear program we define maximizes the per-occupation time gap between the tasks with the least and most single-instance time spent.
We set this objective because we assume that for most occupations there is a large such gap when considering all tasks in O*NET.
A natural alternative which also admits straightforward optimization is to optimize time share weights to minimize distance from the uniform distribution: we state the linear program, derive the associated weights, and show they do not substantially affect our conclusions in \autoref{app:alt_lp_obj}.

In practice, we find the full constraint set is often not feasible because of the $\epsilon$ margin required in per-instance task time between consecutive tasks for the nontrivial values of $\epsilon$ we consider.
Therefore, we iteratively relax the constraints until the problem can be solved by removing the ranking constraints starting with the lowest-ranked task (i.e., the task with the least per-instance time spent). In practice, we executed this fallback, removing the ranking constraints for at least one task, for $64\%$ of occupations. In practice, this also explains why the exact objective of the linear program (e.g., maximizing the gap $w(t^{\text{top}}_o,  o) - w(t^{\text{bot}}_o,  o)$, minimizing the gap from the uniform distribution) has minimal impact, since this means the number of feasible solutions is small in the final iteration of this procedure.

\paragraph{Interpretation.} 
Our factorization reveals a core distinction that is often unaddressed when reasoning about time spent on tasks.
The share of a worker's time spent performing a specific task may be large because the task is performed very frequently (i.e., $\mathbb{E}\left[f(t,o)\right]$ is large) or because each instance of the task is very time consuming (i.e., $w(t, o)$ is large).
This time share is the quantity $s(t, o) = w(t, o)\, \mathbb{E}[f(t, o)]$.
While this share is generally the quantity of interest for most economic questions, our decomposition allows us to focus on validating the $w(t, o)$ values and the associated LM binary preferences, since we assume the correctness of the O*NET frequencies used to compute $\mathbb{E}[f(t, o)]$. 

\textbf{Updating time share estimates~~} Our framework is flexible and naturally accommodates new jobs and change in how jobs are performed. If an occupation's set of tasks changes—for example, tasks are added or removed—or if the frequency at which tasks are performed shifts, the linear program in \autoref{eq:lp} can simply be re-solved with the updated task list or frequency values. And if the per-instance time for a task changes—for example, because workers adopt new AI tools—the prompt we use to elicit an LM ranking of tasks by single-instance time (\autoref{fig:time-preference-prompt}) can be updated to reflect that.

\section{Validation}
\label{sec:val}

Our time shares $s(t, o)$ are the byproduct of a multi-step estimation method.
While we could attempt to recruit workers from different occupations to validate our end-to-end estimates, we anticipated direct validation would make it hard to disambiguate genuine intra-occupation cross-worker heterogeneity in time spent on tasks and undesired noise from flawed estimates.
Therefore, we validate the three components of $s(t, o)$: 
(i) the size of the feasible set of solutions to the constraint satisfaction problem, (ii) the ranking of single-instance time spent per task produced from the language model binary preferences, and (iii) the daily task frequencies $\mathbb{E}[f(t, o)]$ produced from O*NET frequencies.
We use quantitative methods to validate (i), human studies to validate (ii), and spot check (iii) since we generally assume the correctness of O*NET throughout this work.
Together, this provides the basis for claiming the validity of the 
$s(t, o)$ estimates. 

\textbf{The feasible set of the linear program in \autoref{eq:lp} is small.~~} To assess whether the frequency and ranking constraints in \autoref{eq:lp} pin down $w(t, o)$ tightly enough, we measure how far any alternative feasible solution can deviate from the maximin solution we report. For each occupation $o$ we drop the maximin objective and retain only the feasible polytope defined by the constraints in \autoref{eq:lp}. We then probe this polytope along a large set of linear objectives: for every task $t$ we solve LPs that separately maximize and minimize $w(t, o)$, and we additionally sample $1{,}000$ random unit directions ($500$ from a Rademacher distribution and $500$ from a Gaussian distribution) and solve LPs in both the $+d$ and $-d$ directions. Each resulting vertex $w'$ is mapped to its per-task daily time $s'(t, o) = w'(t, o)\,\mathbb{E}[f(t, o)]$ and compared against our reported $s(t, o)$.

Across $876$ occupations, the largest $\ell_2$ deviation across the full task vector, maximized over all sampled alternatives), averages $1.50$ (std $1.20$; max $6.05$; min ${\approx}\,6{\times}10^{-5}$). Against a $7$-hour workday, this indicates the feasible region is concentrated around our reported solution: the most different sampled weight vectors within the feasible region differ by an average of 1.5 hours in weight across all tasks. 

\textbf{Human workers induce a correlated ranking over task time.~~}We recruit full-time workers from five occupations: human-resource managers, lawyers, secretaries and administrative assistants, software developers, and customer service representatives. These occupations were selected to span a diverse set of work types while remaining practical to recruit from.
Each subject completes a three-stage study. First, they are trained on what constitutes a \emph{single instance} of a task. Second, they label pairs of tasks within their occupation by indicating which task a typical worker would spend more time on per instance. Third, we aggregate their pairwise labels into a personal ranking via Copeland's method, show the ranking back to the subject, and allow them to edit it before submission. We filter out subjects who provide low-quality data or fail comprehension checks, as detailed in \autoref{app:human_study}.
After filtering, we retain 10--11 workers per occupation. 

We aggregate individual human rankings into a single human-induced ranking per occupation using Copeland's method, and compare it to the ranking implied by our LP solution in \autoref{eq:lp} via Kendall's $\tau$.
\autoref{tab:human_validation} reports the results, where we also show the mean correlation between the Copeland-aggregated human rankings and each individual human's rankings. Across all five occupations, the LP-implied ranking is positively correlated with the human-induced ranking ($\tau$ ranging from $0.50$ to $0.66$), with four of five correlations significant at the $p < 0.05$ level. The remaining occupation, customer service representatives, has $\tau = 0.50$ ($p = 0.061$), just above the conventional significance threshold. 

Moreover, across all 5 occupations, the LP-implied ranking agrees with the Copeland-aggregated human ranking about as well as the typical individual worker does, when measured against the mean correlation between each worker's ranking and the aggregate. This suggests the gap between our ranking and the human consensus is comparable to the gap between any single worker and that consensus—i.e., our method's disagreement with the aggregate is on the order of the disagreement already present among workers themselves.

These results provide evidence that our LP-produced rankings track how human workers themselves rank tasks by single-instance time. 
One limitation of this analysis, and of our time estimates, is that for many of these occupations there likely exists substantial intra-occupation heterogeneity in time spent per task. 
Regardless, our single-instance time rankings are significantly correlated with the aggregated human worker rankings in $4/5$ occupations. 

\textbf{Task frequencies pass a spot check.~~}
Throughout this work, we assume the correctness of O*NET, including for task frequencies. 
However, to confirm the face validity of these values, we briefly assess individual values.
For a sample of occupations, we examined the least and most frequent tasks to see if they aligned with our expectations for what occupies a work day.
For dental hygienists, cleaning deposits from teeth ($6.01$/day) has a higher expected daily task frequency than running community health clinics ($0.001$/day); for airline pilots, continuous instrument monitoring ($5.26$/day) far exceeds occasionally evaluating other pilots for proficiency ($0.43$/day); and for oral and maxillofacial surgeons, administering anesthesia ($4.63$/day) has a greater daily task frequency than laser surgery for snoring problems ($0.084$/day).


\begin{table}[h]
  \centering
  \begin{tabular}{lccc}
  \toprule
  Occupation & $N$ & Kendall's $\tau$ (LP vs.\ agg.) & Mean Kendall's $\tau$ (agg. vs.\ indv.) [frac.\ sig.] \\
  \midrule
  Human resources managers                  & 10 & $0.66^{**}$ & $0.50$ [$0.60$] \\
  Lawyers                                    & 10 & $0.58^{*}$  & $0.67$ [$0.90$] \\
  Secretaries and administrative assts.  & 10 & $0.61^{*}$  & $0.65$ [$0.80$] \\
  Software developers                        & 11 & $0.57^{*}$  & $0.56$ [$0.64$] \\
  Customer service representatives           & 11 & $0.50$      & $0.46$ [$0.55$] \\
  \bottomrule
  \end{tabular}
  \vspace{1mm}
  \caption{\textbf{Agreement between our per-instance task rankings and human rankings.}
  Kendall's $\tau$ (LP vs.\ agg.) is the Kendall's $\tau$ correlation between the ranking implied by solving our linear program and the Copeland-aggregated human ranking. Mean Kendall's $\tau$ (agg. vs.\ indv.) is the mean Kendall's
  $\tau$ between the aggregate human ranking and each individual worker's ranking; the bracketed value is the fraction of workers whose individual ranking is significantly correlated ($p<0.05$) with the aggregate.
  $N$ is the number of workers per occupation after filtering. ${}^{*} p < 0.05$, ${}^{**} p < 0.01$.}
  \label{tab:human_validation}
  \end{table}

\section{Analysis}
\label{sec:analysis}

We compute $s(t|o)$ and perform all related analysis with data from O*NET 30.2. Constructing our framework's LM ranking cost \$250 with GPT-5.2 across 876 occupations, and can therefore be re-run cheaply as the labor market changes.

\begin{figure}[t]
    \centering
    \begin{subfigure}[t]{0.48\textwidth}
        \centering
        \includegraphics[width=\linewidth]{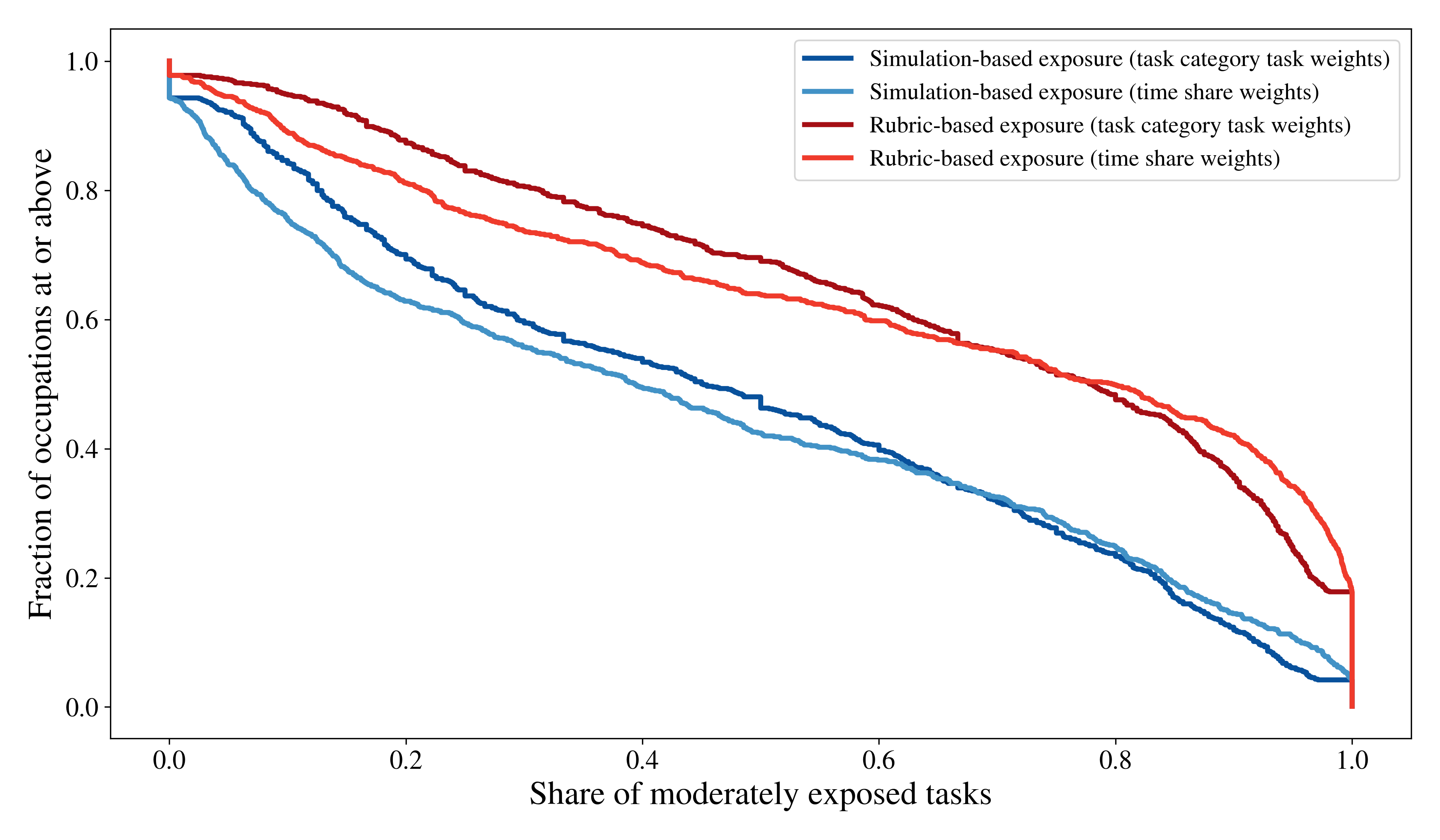}
        \caption{Moderate exposure.}
        \label{fig:exposure-moderate}
    \end{subfigure}
    \hfill
    \begin{subfigure}[t]{0.48\textwidth}
        \centering
        \includegraphics[width=\linewidth]{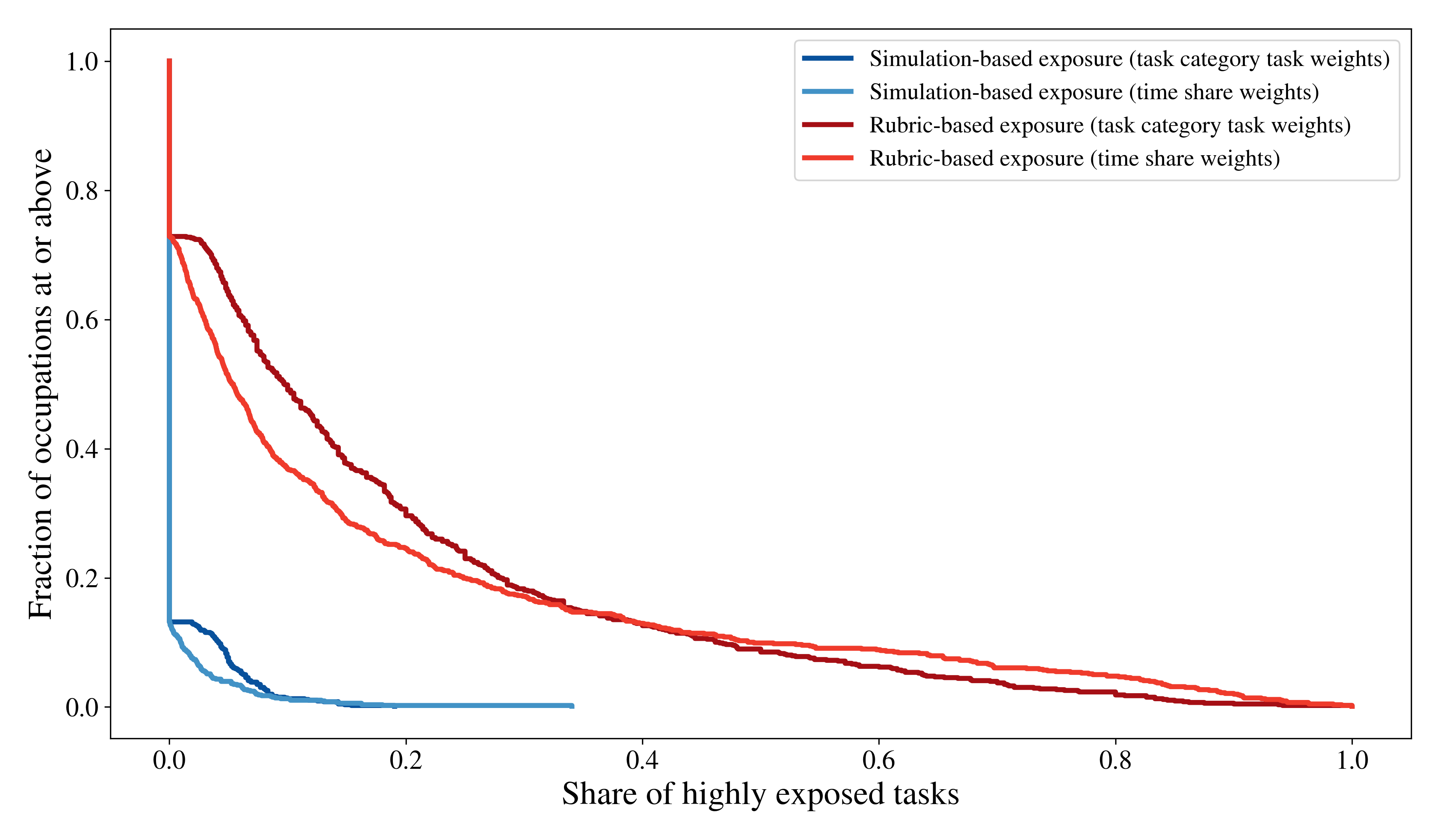}
        \caption{High exposure.}
        \label{fig:exposure-full}
    \end{subfigure}
    \caption{Fraction of occupations at or above a given fraction of tasks exposed to LMs, under rubric-based \citep{eloundou2024gpts} and simulation-based \citep{wan2026econevals} exposure measures. Each panel compares two weighting schemes: weighting tasks by our time-share estimates $s(t, o)$ versus the core/supplemental weighting of \citet{eloundou2024gpts}. Panel~(a) labels a task as exposed if it meets a moderate exposure threshold (e.g., $>25\%$ time saved), while panel~(b) labels a task as exposed if it meets a high exposure threshold (e.g., $>50\%$ time saved).}
    \label{fig:exposure-comparison}
\end{figure}

\textbf{Re-estimating occupational exposure to language models.~~} 
We examine how weighting tasks by their time share affects occupation-level exposure estimates.
We consider two task-level exposure measures: rubric-based exposure \citep{eloundou2024gpts} and simulation-based exposure \citep{wan2026econevals}.
The exposure estimates are published by \citet{wan2026econevals}: this work introduces simulation-based exposure and provides rubric-based alternatives based on the updated exposure rubric \citet{hosseini2026generative_entry_barriers}. 
For task weights, we consider the baseline weights popularized by \citet{eloundou2024gpts}: a weight of 1 for tasks labeled in O*NET as core to an occupation and 0.5 for tasks labeled as supplemental. 
\autoref{fig:exposure-comparison} presents the results, with further plotting details provided in \autoref{app:plotting_details}.

The general trend is time weights decrease  exposure: exposed tasks tend to occupy less worker time. 
However, this pattern reverses for the most exposed occupations. Put together, this suggests that common weighting approaches distort occupational exposure estimates in different ways. Namely, prior estimates that do not account for time shares are likely to overstate the time saved for most occupations while understating the time saved for the occupations with the greatest time savings.

\begin{figure}[h]
\centering
\includegraphics[width=0.95\linewidth]{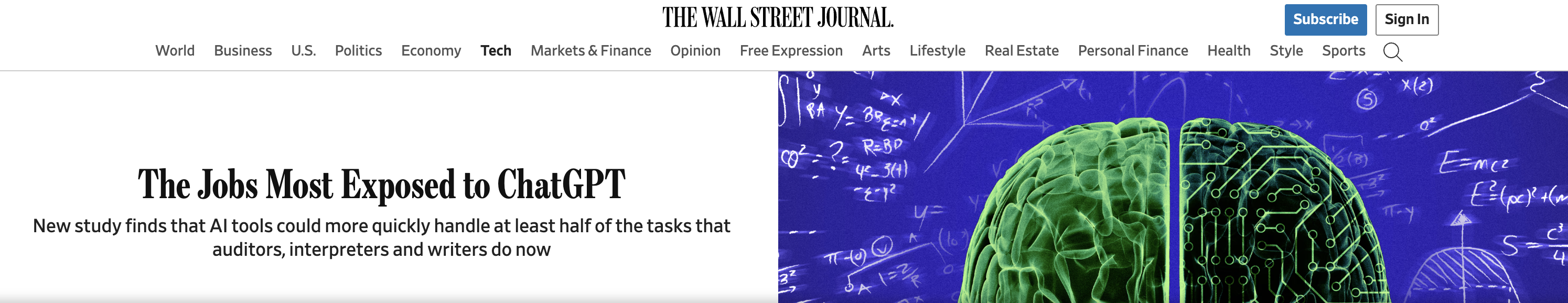}
\caption{Source: \citet{weber2023wsj}, \textit{The Wall Street Journal}. The jobs that have the highest exposure as measured by the rubric-based exposure measure from \cite{eloundou2024gpts} are widely reported in press.}
\label{fig:wsj}
\end{figure}

\textbf{Accounting for time share changes among the top 25 most exposed occupations.~~} 
Government, industry, and the public are all intensely concerned with how AI will affect jobs.
Therefore, research on the economic impacts of AI is of broad interest and often features directly in news.
The work of \citet{eloundou2024gpts} has been widely covered because it provides a straightforward apparatus for connecting technological capabilities to jobs \citep{weber2023wsj, euronews2023openai, pcmag2023chatgpt, vice2023openai, decoder2023gpt4,
businesstoday2023altman, bandt2023openai}.
For example, the Wall Street Journal published the headline ``These are the jobs most affected by AI'' (Figure \ref{fig:wsj}) \citep{weber2023wsj}. 
And since the most exposed occupations may be attractive targets for policies to govern AI's labor impacts, we observe a similar through-line from AI exposure research to government reports \citep[e.g.,][]{scavette2025occupational, gmyrek2024genai}.
For example, the most exposed occupations feature in reports prepared by U.S. Senator Bernie Sanders as Ranking Member of the U.S. Senate Committee on Health, Education, Labor and Pension.  
Because these government and policy reports aim to inform real policy decisions, the weighting used to rank exposure matters: each relies on the share of a job's O*NET tasks that are exposed, rather than the share of working time exposed to AI.

Having established that occupational exposure estimates are sensitive to time weights, we consider how this sensitivity would change the top-level findings from prior occupational exposure research.
Since most media and policy coverage of prior work focuses on the most-exposed occupations, we consider the top-25 occupations using rubric-based task exposure estimates based on (i) core/supplemental weights vs. (ii) our time weights.
Changes to the most-exposed occupations could alter media narratives and policymaker priorities on AI's labor impacts.

Time weights substantially modify the set of occupations that are most exposed (\autoref{tab:hosseini_high_top25_swaps}).\footnote{However, the bottom-25 occupations do not change when re-weighting by time.} 
11 occupations originally in the top 25 drop out with 9 being clerical or customer service occupations.\footnote{The other two occupations are Web Administrators and Information Security Engineers. Both occupations have unexposed planning and collaboration tasks that take significant time.}
We find that the cause for this shift is that time weights de-emphasize the time savings for transactional tasks (e.g., updating a record, processing a claim, or answering a customer inquiry) because the overall time spent on these tasks is limited.
In particular, beyond these transactional tasks, a small handful of other tasks that involve judgment or interpersonal interaction occupy more of a worker's time and are not very exposed to current AI capabilities.
The time-weighted top-25 list includes several analytical occupations that have only one or two tasks that are exposed.
These research, drafting, or modeling tasks consume a lot of time even though they are dominated by quick administrative tasks in the task count for these occupations.

We underscore that an occupation that has highly exposed time share may not necessarily see a decline in wages or employment, nor is it entirely automatable; the effects of exposure on employment and wages is an open research direction and dependent on many other factors, such as task expertise \citep{autor2025expertise} or the structure of bottlenecks \citep{gans2026oring}. 

\begin{table}[h]
\centering
\small
\begin{tabular}{p{0.46\linewidth} p{0.46\linewidth}}
\toprule
\textbf{Left the top-25 most exposed occupations when weighting by time share} & \textbf{Entered the top-25 most exposed occupations when weighting by time share} \\
\midrule
Online Merchants & Training and Development Managers \\
Web Administrators & Credit Analysts \\
Information Security Engineers & Financial Quantitative Analysts \\
Travel Agents & Computer Programmers \\
Telemarketers & Geographic Information Systems Technologists \\
Bookkeeping, Accounting, and Auditing Clerks & Actuaries \\
Brokerage Clerks & Statisticians \\
Customer Service Representatives & Economists \\
Loan Interviewers and Clerks & Judicial Law Clerks \\
Desktop Publishers & Poets, Lyricists, and Creative Writers \\
Insurance Claims and Policy Processing Clerks & Credit Authorizers, Checkers, and Clerks \\
\bottomrule
\end{tabular}
\caption{Occupations that swap in and out of the top-25 most exposed under rubric-based exposure measure when switching from task-category weighting to our proposed time-share weighting. $11/25$ occupations swap out.}
\label{tab:hosseini_high_top25_swaps}
\end{table}

\textbf{Examples of occupational exposure differences.~~}Figure \ref{fig:massage-exposure} shows an example of how re-weighting tasks by time paints a different picture of occupational exposure for Massage Therapists. We discuss two additional contrasting occupations that illustrate the mechanism behind this divergence under the rubric-based exposure measure. For Forest and Conservation Workers (O*NET 45-4011.00), $29\%$ of tasks are labeled highly exposed. However, the most time-intensive tasks---operating machinery like bulldozers, cutting trees, identifying diseased trees, maintaining campsites--are not flagged exposed; the exposed tasks---maintaining tallies of trees examined, checking equipment or regulations---each carry small time weights. Weighting by our time-share estimates, we estimate that $\sim 4\%$ of a Forest and Conservation Worker's working time is exposed to AI. For Actuaries (O*NET 15-2011.00) the pattern reverses: $37\%$ of tasks are labeled highly exposed, however, when we re-weight by time share, we predict $89\%$ of an Actuaries' time is highly exposed. The tasks that are not highly exposed are those that require interpersonal interaction---negotiating terms and conditions, explaining changes in contract provisions to customer, testifying before public agencies---but we predict most of an Actuaries' time is spent on highly exposed tasks like analyzing statistical information, calculating premiums, or determining financial soundness.

\textbf{Correlation with \cite{tamkinmccrory2025productivity} time estimates.~~} First, we compare our time share estimates, $s(t|o)$, to those of \citet{tamkinmccrory2025productivity}, which are obtained by directly prompting an LM to estimate the share of a worker's time spent on each task. We use the dataset released by \citet{tamkinmccrory2025productivity}, and run a regression test over the 16{,}510 (occupation, task) pairs common to both datasets. We find a strongly positive relationship (slope $\beta = 5.29$ hours/day per unit share, $\text{SE} = 0.07$, $R^2 = 0.237$, $p < 0.01$); adding occupation fixed effects leaves the coefficient nearly unchanged ($\beta = 5.13$, $R^2 = 0.252$), indicating that our shares track theirs not only across occupations but also across tasks within the same occupation. That the slope falls below the 7-hour workday ceiling implies our time shares are somewhat closer to uniform than \citet{tamkinmccrory2025productivity}'s. Additionally, running a Spearman's test results in $\rho = 0.55$ ($p < 0.01$).

Next, we compare our single-instance task time estimates $w(t, o)$ to the human task completion times reported by \citet{tamkinmccrory2025productivity}, which are inferred by an LM from Claude chat logs. Note that these serve as a proxy for task complexity, a quantity distinct from what $w(t, o)$ measures. We find a moderate positive Spearman rank correlation of $\rho = 0.54$ ($p < 0.01$) between $w(t, o)$ and the completion times reported by \citet{tamkinmccrory2025productivity}. Tasks that our method estimates take longer to complete in a single instance also tend to be those whose corresponding Claude conversations take longer for a human to complete end-to-end.

\textbf{Comparison to O*NET descriptors.~~} We compare our per-task time estimates $s(t, o)$ against the three O*NET descriptors discussed in Section~2: per-task importance, per-task frequency, and the binary core/supplemental task-type classification. We find that these descriptors capture related but distinct features from our time share estimates.

We evaluate the correlation between task rankings for each occupation when ranking by mean importance and ranking by $s(t, o)$. The correlation between task rankings is statistically significant ($p < 0.05$) in 504 of 876 occupations, with a mean Kendall's $\tau$ of $0.47$ among occupations with significantly correlated rankings. The composite measure of importance times frequency, restricted to core tasks, yields stronger ranking agreement: rankings are significantly correlated in 575 of 876 occupations, with a mean Kendall's $\tau$ of $0.61$ among occupations with significant rankings. O*NET frequency on its own, however, produces an essentially uncorrelated ranking: only 131 of 876 occupations attain a significantly correlated ranking, and among those the mean correlation is $-0.317$. 
 
 The binary core/supplemental classification, which is used by \cite{eloundou2024gpts} as a substitute for time weights when computing occupational exposure, explains only $1.3\%$ of the variance in $s(t, o)$ when running a regression test over $17{,}525$ tasks across $876$ occupations. Core tasks do have a higher mean predicted time share than supplemental tasks ($0.38$ vs.\ $0.25$ hours per day; $p<0.01$),
 
 

\section{Discussion}

We present a method to compute the share of time spent by workers per task in O*NET, grounded in transparent assumptions and work data collected from O*NET in addition to LLM inferences. This quantity enables us to compute the share of working time exposed to LMs, rather than the share of tasks, offering a different perspective on the impact of LMs on jobs. Across nearly all occupations in the U.S., we find that time-share weighting systematically lowers measured LM exposure for most occupations but raises it for the most exposed ones, and reshuffles 11 of the 25 occupations reported as most exposed — shifting the top of the list away from clerical work and toward analytical roles. More broadly, because our estimates are a set of weights over the tasks within each occupation, they can be substituted directly into any task-based analysis that aggregates to the occupation level.

Beyond what is already discussed, several limitations point to directions for future work. First, the quantity we estimate likely varies substantially across workers within an occupation. While we expect our time-share weights to be more accurate than other coarser weighting schemes, better characterizing the within-occupation heterogeneity could yield more practical estimates. Additionally, the feasible set of solutions satisfying our constraints that define the time share estimates is non-negligible; future work could explore how to further tighten this set, as well as alternative criteria for selecting a solution from it. Finally, while we apply our time-share weights to occupational exposure, we expect that other task-based analyses would similarly benefit from incorporating them.

\section*{Acknowledgements}
We thank Abhishek Nagaraj, Amir Zeinali, Andy Haupt, Arjun Ramani, Arul Murugan, Avanika Narayan, Bharat Chandar, Connacher Murphy, Diyi Yang, Dylan Clement, Erik Brynjolfsson,  Joel Becker, Jon Saad-Falcon, Kris Gulati, Lukas Freund, Lukas Mann, Phil Trammell, Parker Whitfill, Sam Manning, Tom Cunningham, and Yijia Shao for helpful discussion.
We thank the Stanford Center for Research on Foundation Models (Stanford CRFM) and Stanford Institute for Human-Centered Artificial Intelligence (Stanford HAI) for funding.

\bibliographystyle{plainnat}   
\bibliography{references}   

\newpage
\appendix

\section{Human Study Details}
\label{app:human_study}
We recruit subjects for our IRB approved study (Protocol Number: IRB-85779) from Prolific.com. We mandate that all workers are residents of the United States, are employed full time in the occupation we are requesting data for (self-reported), and have an approval rating of at least $80\%$. An example of the study is available online.\footnote{\url{https://main.d1kt072szdxujv.amplifyapp.com/15-1252.00}}. We pay workers $\$15$ for completing the study, which we estimate is on the order of $\$25$ per hour. 

To ensure high quality responses, we additionally remove worker data using the following criteria:
\begin{itemize}
    \item We have an attention check where we ask people to select a specific option, and remove the data of workers who do not follow that instruction.
    \item Of the pairwise comparisons we ask each worker to make, $5$ are presented twice: once in their original orientation and once with the two tasks swapped. We remove workers who give inconsistent verdicts (e.g., select Task 1 in one orientation and Task 2 in the reversed one) on more than $2$ of these $5$ twin pairs.
    \item Before the main study, each worker completes $8$ practice pairwise comparisons that have an objectively correct answer (e.g., comparing a clearly short task against a clearly long one). We remove workers who answer fewer than $7$ of these $8$ practice questions correctly.
\end{itemize}

\autoref{tab:all_workers_per_cond} shows the number of participants who we collected data from, and the number of participants who passed our filter. We paid a total of $\$1,290$ to participants to collect this data.

\begin{table}[t]
  \centering
  \begin{tabular}{lcc}
  \toprule
  Occupation & \begin{tabular}{@{}c@{}}Number of participants who\\passed comprehension checks\end{tabular} & Total participants \\
  \midrule
  Software Developers & 11 & 14 \\
  Customer Service Representatives & 11 & 18 \\
  Lawyers & 10 & 16 \\
  Human Resources Managers & 10 & 15 \\
  Secretaries and Administrative Assistants & 10 & 23 \\
  \bottomrule
  \end{tabular}
  \caption{Participants collected and passing comprehension checks per occupation.}
  \label{tab:all_workers_per_cond}
  \end{table}

\section{Time Share Method details}
\label{app:method_details}

\subsection{Establishing the confidence interval}
When prompting an LLM to label pairs of tasks to rate which one's single-task time instance is greater, we remove all labels that have a probability-of-generation of less than $70\%$. We chose this value by iterating over the set of thresholds $[60\%,70\%,80\%,90\%]$ and choosing the value that resulted in both the highest Spearman's correlation with the time estimates released by \cite{tamkinmccrory2025productivity} and the smallest feasible region as measured using the procedure discussed in \autoref{sec:val}. 

\section{Alternative LP objectives}
\label{app:alt_lp_obj}
Here we explore an alternative objective for the LP in \autoref{eq:lp}; finding the most uniform set of time share weights that satisfy the set of constraints. More formally, we solve the following LP:

\begin{equation}
\label{eq:lp_alt}
\begin{aligned}
\max_{\{w(t, o)\},\, \{d_o\}} \quad & \sum_{o} d_o \\
\text{s.t.} \quad
& \sum_{t \,\in\, \mathrm{tasks}(o)} w(t, o)\, \mathbb{E}[f(t, o)] \;=\; 7, && \forall\, o, \\
& w(t_i, o) \;\ge\; w(t_j, o) + \epsilon, && \forall\, (t_i, t_j) \in \mathcal{R}_o,\; \forall\, o, \\
& d_o \;\le\; w(t, o)\, \mathbb{E}[f(t, o)], && \forall\, t \in \mathrm{tasks}(o),\; \forall\, o, \\
& w(t, o) \;\ge\; 0, && \forall\, t, o.
\end{aligned}
\end{equation}

\begin{figure}[t]
    \centering
    \begin{subfigure}[t]{0.48\textwidth}
        \centering
        \includegraphics[width=\linewidth]{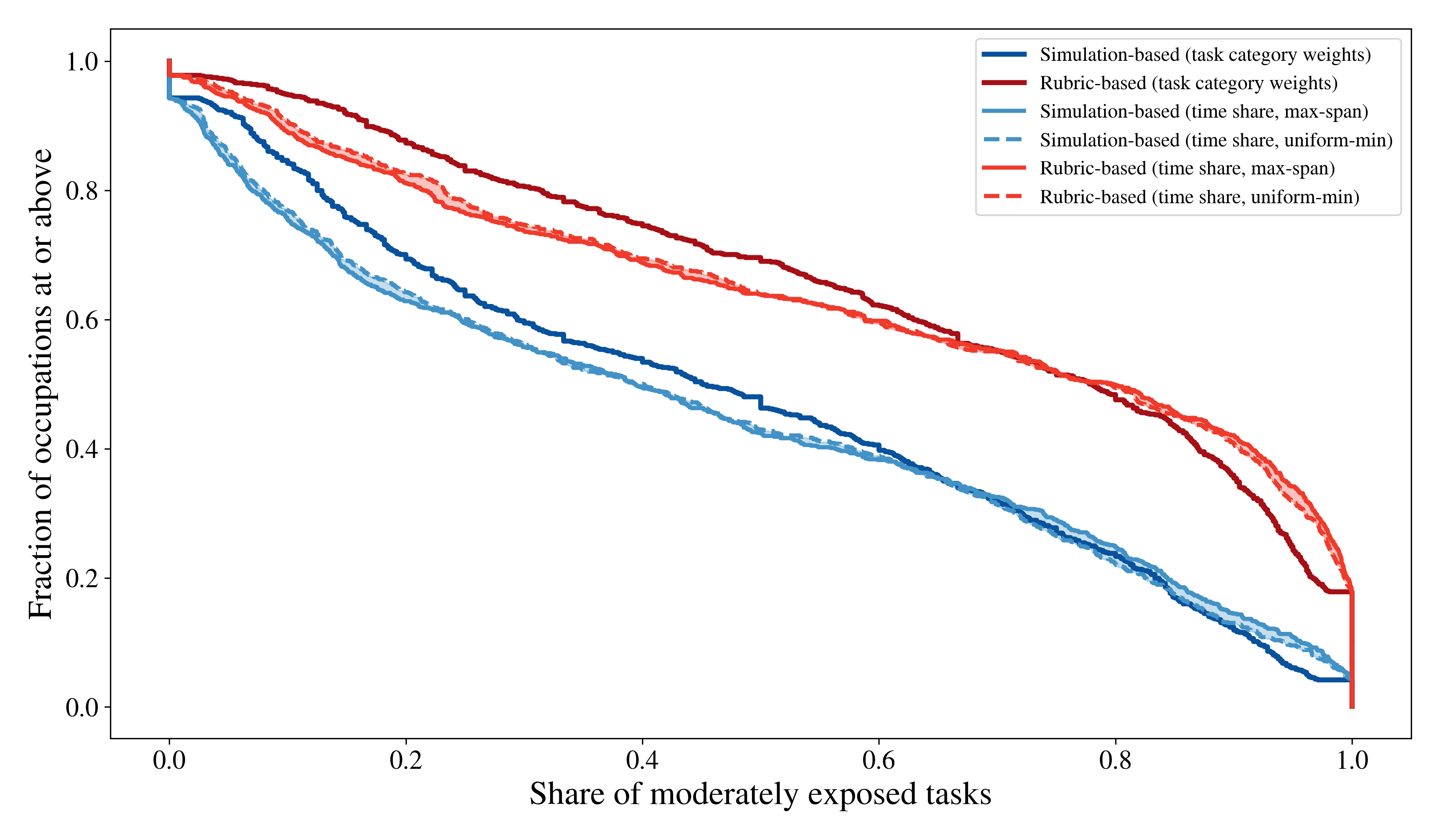}
        \caption{Moderate exposure.}
        \label{fig:exposure-moderate}
    \end{subfigure}
    \hfill
    \begin{subfigure}[t]{0.48\textwidth}
        \centering
        \includegraphics[width=\linewidth]{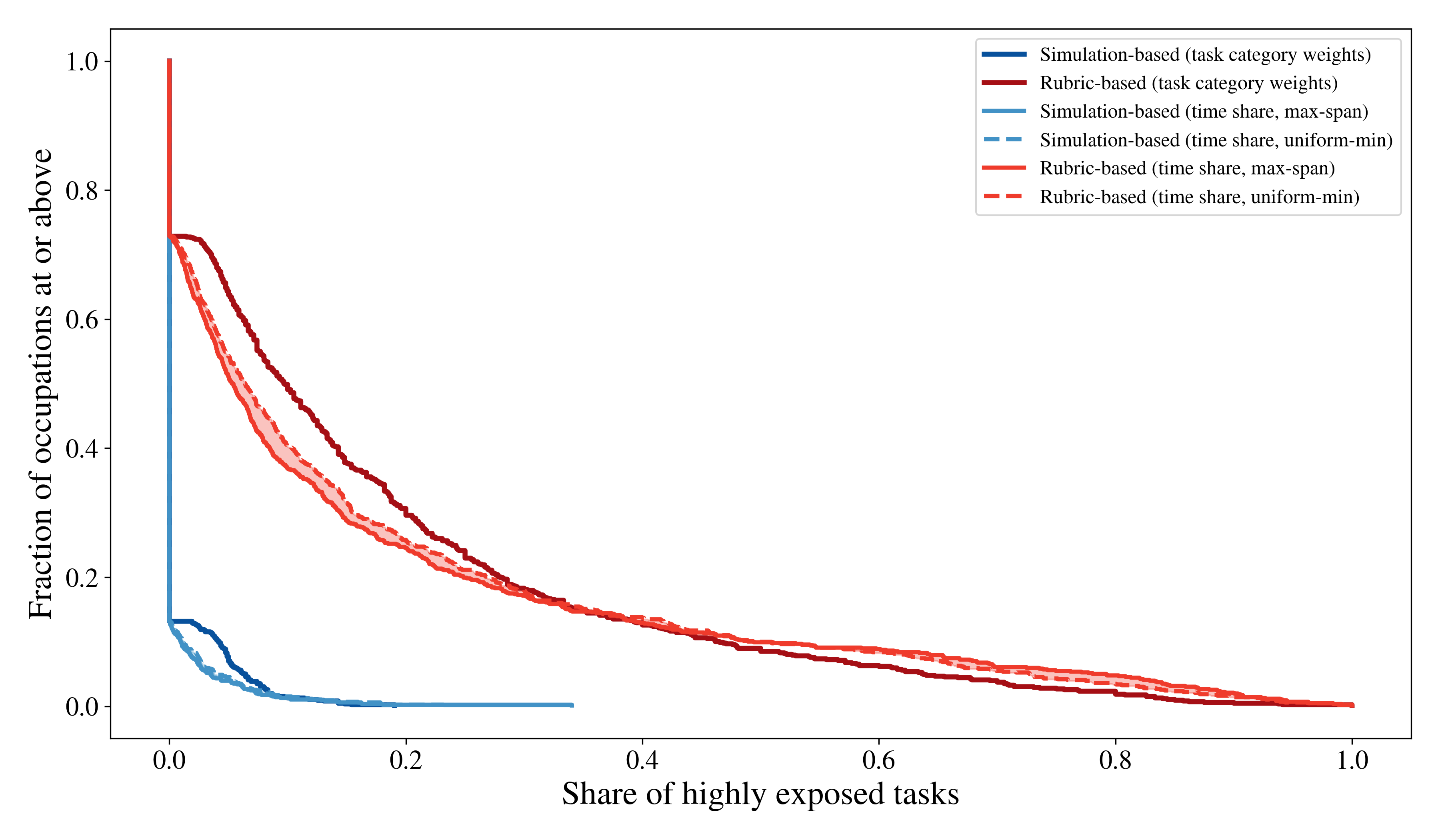}
        \caption{High exposure.}
        \label{fig:exposure-full}
    \end{subfigure}
    \caption{Fraction of occupations at or above a given fraction of tasks exposed to LMs, under rubric-based \citep{eloundou2024gpts} and simulation-based \citep{wan2026econevals} exposure measures. Each panel compares two weighting schemes: weighting tasks by our time-share estimates $s(t, o)$ versus the core/supplemental weighting of \citet{eloundou2024gpts}. We additionally plot two different time-share estimates; the solid line uses the time share estimates found via \autoref{eq:lp} and the dotted line via \autoref{eq:lp_alt}, with the difference between the two shaded in. Panel~(a) labels a task as exposed if it meets a moderate exposure threshold (e.g., $>25\%$ time saved), while panel~(b) labels a task as exposed if it meets a high exposure threshold (e.g., $>50\%$ time saved).}
    \label{fig:exposure-comparison-ablation}
\end{figure}

As a complement to \autoref{fig:exposure-comparison}, in \autoref{fig:exposure-comparison-ablation} we plot the exposure intensity computed using the time weights from \autoref{eq:lp_alt} and from \autoref{eq:lp}, shading the area between the two curves. The choice of LP objective does not have a substantial effect on the resulting share of time exposed under any considered exposure measure.

\section{Independence Assumption in Time Share Factorization}
\label{app:ind_assumption}

Our time share estimate $s(t, o) = w(t, o) \cdot \mathbb{E}[f(t, o)]$ treats
per-instance time $w(t, o)$ and expected task frequency $\mathbb{E}[f(t, o)]$
as independent quantities. If the expectation is taken over workers within an
occupation, this factorization requires that $w$ and $f$ are uncorrelated across
workers, i.e., $\mathbb{E}[w(t,o) \cdot f(t,o)] = \mathbb{E}[w(t,o)] \cdot
\mathbb{E}[f(t,o)]$. This assumption may be violated: for example, workers who
complete a task instance faster may take on more instances, inducing a negative
within-occupation correlation between $w$ and $f$. In this case,
$\mathbb{E}[w \cdot f] < \mathbb{E}[w] \cdot \mathbb{E}[f]$, and our estimates
would overstate the expected time share.

We acknowledge this as a limitation of our factorization. The direction and magnitude of any such correlation is empirically ambiguous: while faster workers may perform more instances of a task, occupational norms, workflow structure, and task assignment practices often constrain individual task frequency independently of individual speed. Resolving this would require individual-level data linking task duration to task frequency within occupations, which is unavailable at the scale of O*NET's coverage.

\section{Full Top-25 Lists Under Both Weighting Schemes}
\label{app:top25-full}

To complement Table~\ref{tab:hosseini_high_top25_swaps}, which reports only the occupations that enter and exit the top 25, we present the complete top-25 most-exposed occupation lists under both weighting schemes in Table~\ref{tab:top25-taskcat} and \ref{tab:top25-timeshare}

\begin{table}[t]
\centering
\small
\begin{tabular}{rllcc}
\toprule
Rank & O*NET-SOC & Occupation & Task-category & Time-share \\
\midrule
1  & 15-1243.01 & Data Warehousing Specialists$^\dagger$            & 1.000 & 1.000 \\
2  & 15-2051.01 & Business Intelligence Analysts$^\dagger$          & 0.941 & 0.991 \\
3  & 43-9022.00 & Word Processors and Typists$^\dagger$             & 0.938 & 0.939 \\
4  & 41-9041.00 & Telemarketers                                     & 0.900 & 0.859 \\
5  & 43-3051.00 & Payroll and Timekeeping Clerks$^\dagger$          & 0.875 & 0.900 \\
6  & 15-2099.01 & Bioinformatics Technicians$^\dagger$              & 0.862 & 0.963 \\
7  & 41-3041.00 & Travel Agents                                     & 0.857 & 0.823 \\
8  & 43-9081.00 & Proofreaders and Copy Markers$^\dagger$           & 0.850 & 0.945 \\
9  & 43-4131.00 & Loan Interviewers and Clerks                      & 0.844 & 0.792 \\
10 & 15-1254.00 & Web Developers$^\dagger$                          & 0.840 & 0.984 \\
11 & 13-2082.00 & Tax Preparers$^\dagger$                           & 0.833 & 0.923 \\
12 & 43-4051.00 & Customer Service Representatives                  & 0.833 & 0.781 \\
13 & 43-4021.00 & Correspondence Clerks$^\dagger$                   & 0.826 & 0.870 \\
14 & 43-9111.00 & Statistical Assistants$^\dagger$                  & 0.826 & 0.886 \\
15 & 15-1253.00 & Software QA Analysts and Testers$^\dagger$        & 0.808 & 0.905 \\
16 & 31-9094.00 & Medical Transcriptionists$^\dagger$               & 0.800 & 0.927 \\
17 & 43-4011.00 & Brokerage Clerks                                  & 0.800 & 0.833 \\
18 & 43-9031.00 & Desktop Publishers                                & 0.800 & 0.839 \\
19 & 43-9041.00 & Insurance Claims and Policy Processing Clerks     & 0.800 & 0.679 \\
20 & 43-9021.00 & Data Entry Keyers$^\dagger$                       & 0.769 & 0.941 \\
21 & 15-1299.01 & Web Administrators                                & 0.768 & 0.601 \\
22 & 43-3031.00 & Bookkeeping, Accounting, and Auditing Clerks      & 0.757 & 0.803 \\
23 & 15-1299.05 & Information Security Engineers                    & 0.750 & 0.821 \\
24 & 13-1081.02 & Logistics Analysts$^\dagger$                      & 0.736 & 0.886 \\
25 & 13-1199.06 & Online Merchants                                  & 0.731 & 0.683 \\
\bottomrule
\end{tabular}
\caption{Top-25 most exposed occupations under core/supplemental task-category weighting (rubric-based), sorted by task-category exposure score. Occupations marked $\dagger$ appear in both top-25 lists.}
\label{tab:top25-taskcat}
\end{table}

\begin{table}[t]
\centering
\small
\begin{tabular}{rllcc}
\toprule
Rank & O*NET-SOC & Occupation & Task-category & Time-share \\
\midrule
1  & 15-1243.01 & Data Warehousing Specialists$^\dagger$            & 1.000 & 1.000 \\
2  & 15-2051.01 & Business Intelligence Analysts$^\dagger$          & 0.941 & 0.991 \\
3  & 15-1254.00 & Web Developers$^\dagger$                          & 0.840 & 0.984 \\
4  & 13-2099.01 & Financial Quantitative Analysts                   & 0.647 & 0.964 \\
5  & 15-2099.01 & Bioinformatics Technicians$^\dagger$              & 0.862 & 0.963 \\
6  & 43-9081.00 & Proofreaders and Copy Markers$^\dagger$           & 0.850 & 0.945 \\
7  & 15-1299.02 & Geographic Information Systems Technologists       & 0.706 & 0.945 \\
8  & 43-9021.00 & Data Entry Keyers$^\dagger$                       & 0.769 & 0.941 \\
9  & 43-9022.00 & Word Processors and Typists$^\dagger$             & 0.938 & 0.939 \\
10 & 31-9094.00 & Medical Transcriptionists$^\dagger$               & 0.800 & 0.927 \\
11 & 13-2082.00 & Tax Preparers$^\dagger$                           & 0.833 & 0.923 \\
12 & 15-1251.00 & Computer Programmers                               & 0.700 & 0.909 \\
13 & 27-3043.05 & Poets, Lyricists, and Creative Writers             & 0.577 & 0.908 \\
14 & 23-1012.00 & Judicial Law Clerks                                & 0.393 & 0.908 \\
15 & 15-1253.00 & Software QA Analysts and Testers$^\dagger$        & 0.808 & 0.905 \\
16 & 15-2041.00 & Statisticians                                     & 0.667 & 0.904 \\
17 & 43-3051.00 & Payroll and Timekeeping Clerks$^\dagger$          & 0.875 & 0.900 \\
18 & 11-3131.00 & Training and Development Managers                  & 0.348 & 0.893 \\
19 & 15-2011.00 & Actuaries                                          & 0.370 & 0.889 \\
20 & 13-1081.02 & Logistics Analysts$^\dagger$                      & 0.736 & 0.886 \\
21 & 43-9111.00 & Statistical Assistants$^\dagger$                  & 0.826 & 0.886 \\
22 & 19-3011.00 & Economists                                         & 0.444 & 0.879 \\
23 & 43-4021.00 & Correspondence Clerks$^\dagger$                   & 0.826 & 0.870 \\
24 & 43-4041.00 & Credit Authorizers, Checkers, and Clerks           & 0.714 & 0.870 \\
25 & 13-2041.00 & Credit Analysts                                    & 0.611 & 0.868 \\
\bottomrule
\end{tabular}
\caption{Top-25 most exposed occupations under time-share weighting (rubric-based), sorted by time-share exposure score. Occupations marked $\dagger$ appear in both top-25 lists.}
\label{tab:top25-timeshare}
\end{table}


\section{Plotting details}
\label{app:plotting_details}

The simulation-based exposure measures produced by \citet{wan2026econevals} are incomplete, missing data for $4\%$ of tasks in O*NET 30.2. 
In \autoref{fig:exposure-comparison} we therefore remove occupations that have less than $80\%$ of their tasks covered by the simulation-based exposure measures, resulting in data for  $858$ occupations. We remove all tasks that do not have simulation-based exposure measures from both the rubric-based and simulation-based occupational exposure measures for a fair comparison, i.e., both exposure measures operate over the same set of tasks. 

\section{Additional experiments}

\subsection{Weighting by time doesn't meaningfully change what skills explain occupational exposure}

To check whether weighting tasks by predicted time substantively changes the occupation-level relationship between O*NET skills and exposure, we re-run the regression used by \citet{eloundou2024gpts}, regressing each occupation's exposure share on its O*NET basic-skills vector. We use two different task weighting schemes: the task-category weighting of \citet{eloundou2024gpts} (Core $= 1.0$, Supplemental $= 0.5$) and our proposed time share weights $s(t|o)$. The regression is run on the $n = 858$ occupations for which both O*NET skill ratings and a complete set of time share estimates are available. \autoref{tab:skill_exposure_time_weighting} reports $R^2$ and adjusted $R^2$ under both rubric-based and simulation-based exposure measures. Adding time weights reduces $R^2$ marginally for both exposure measures and leaves the ranking of skill correlations essentially unchanged.

\begin{table}[h]
\centering
\small
\begin{tabular}{lcccc}
\toprule
 & \multicolumn{2}{c}{Simulation-based exposure} & \multicolumn{2}{c}{Rubric-based exposure} \\
\cmidrule(lr){2-3} \cmidrule(lr){4-5}
Skill & Task category weighting & Time share weighting & Task category weighting &  Time share weighting \\
\midrule
Reading Comprehension & $+0.749$ & $+0.732$ & $+0.779$ & $+0.757$ \\
Writing               & $+0.754$ & $+0.723$ & $+0.760$ & $+0.734$ \\
Speaking              & $+0.668$ & $+0.651$ & $+0.673$ & $+0.660$ \\
Active Listening      & $+0.649$ & $+0.630$ & $+0.687$ & $+0.670$ \\
Critical Thinking     & $+0.665$ & $+0.635$ & $+0.630$ & $+0.617$ \\
Active Learning       & $+0.660$ & $+0.639$ & $+0.621$ & $+0.606$ \\
Programming           & $+0.551$ & $+0.537$ & $+0.568$ & $+0.559$ \\
Learning Strategies   & $+0.553$ & $+0.527$ & $+0.526$ & $+0.517$ \\
Mathematics           & $+0.519$ & $+0.508$ & $+0.520$ & $+0.504$ \\
Monitoring            & $+0.356$ & $+0.338$ & $+0.372$ & $+0.360$ \\
Science               & $+0.279$ & $+0.268$ & $+0.326$ & $+0.316$ \\
\midrule
Overall $R^2$         & $0.721$  & $0.684$  & $0.736$  & $0.697$  \\
\bottomrule
\end{tabular}
\caption{Pearson correlation of each O*NET basic skill with occupation-level exposure share, under task-category vs. time share weighting schemes across $858$ occupations.}
\label{tab:skill_exposure_time_weighting}
\end{table}

\begin{figure*}[t]
\centering
\begin{tcolorbox}[
    enhanced,
    colback=gray!3,
    colframe=black!75,
    boxrule=0.5pt,
    arc=2pt,
    left=8pt, right=8pt, top=6pt, bottom=6pt,
    fonttitle=\bfseries,
    title=Pairwise Task-Time Annotation Prompt,
    coltitle=white,
    colbacktitle=black!70,
    attach boxed title to top left={xshift=6pt, yshift=-2pt},
    boxed title style={colframe=black!75, arc=1pt},
    fontupper=\small\ttfamily,
]
You are a careful annotation model. You are given two O*NET tasks and you need to classify which \emph{single-instance duration} of the tasks is typically more time-consuming for a typical worker in this occupation, under normal working conditions.

\medskip
You will be given:\\
(1)~an O*NET occupation title\\
(2)~two O*NET task statements.

\medskip
\textbf{Your job:} Classify which \emph{single-instance duration} of the tasks is typically more time-consuming for a typical worker in this occupation, under normal working conditions.

\medskip
\textbf{IMPORTANT RULES:}
\begin{enumerate}[leftmargin=*, itemsep=2pt, topsep=2pt]
    \item Disregard frequency entirely. Do NOT use how often the task is performed (daily/weekly/rarely) to infer duration.
    \item Classify only the duration of one completion of the task as written (one run-through).
    \begin{itemize}[leftmargin=*, itemsep=1pt, topsep=1pt]
        \item \emph{Example:} If the task is cleaning a home, a single instance is cleaning a home once.
        \item \emph{Example:} If the task takes multiple sessions to complete (e.g., writing a book), a single instance is the time a typical worker spends to make meaningful progress in a day.
        \item \emph{Example:} If the task requires short, frequent instances (e.g., monitoring employees, answering phone calls), a single instance is one run-through.
    \end{itemize}
    \item Exclude waiting/idle time unless the person must actively monitor or be continuously engaged.
    \item Assume the typical worker performing this task does not have access to Generative AI assistance.
\end{enumerate}

\medskip
\textbf{OUTPUT:} Output only a single label and nothing else:
\begin{itemize}[leftmargin=*, itemsep=1pt, topsep=2pt]
    \item \textbf{0} -- a single instance of task~1 is typically more time-consuming
    \item \textbf{1} -- a single instance of task~2 is typically more time-consuming
    \item \textbf{2} -- the tasks are typically roughly equally time-consuming
    \item \textbf{3} -- you cannot decide (also use 3 if more context is needed or the label is context-dependent)
\end{itemize}

\medskip
\textbf{EXAMPLES:}

\medskip
Occupation: Barber\\
Task 1: Cutting hair\\
Task 2: Answering phone calls\\
Label: \textbf{0}\\
\emph{Explanation:} A single instance of cutting hair typically takes longer than answering one phone call.

\medskip
Occupation: Barber\\
Task 1: Managing payment from customers\\
Task 2: Researching new hairstyles\\
Label: \textbf{1}\\
\emph{Explanation:} Taking one payment typically takes less time than one research session.

\medskip
Occupation: Barber\\
Task 1: Managing payment from customers\\
Task 2: Answering booking phone-calls\\
Label: \textbf{3}\\
\emph{Explanation:} It is not clear which single instance is more time-consuming, so you cannot decide.

\medskip
\textbf{INPUT:}\\
Occupation: \{\textit{occupation\_title}\} (O*NET-SOC: \{\textit{onet\_soc\_code}\})\\
Task 1: \{\textit{task\_description\_1}\}\\
Task 2: \{\textit{task\_description\_2}\}

\medskip
Output exactly one label and nothing else: 0, 1, 2, or 3 (as defined above).
\end{tcolorbox}
\caption{Prompt used to elicit pairwise single-instance time judgments between two O*NET tasks within an occupation.}
\label{fig:time-preference-prompt}
\end{figure*}

\end{document}